\documentclass[hypertexnames=false,twocolumn,aps,pra]{revtex4-2}
\usepackage{array, makecell}

\usepackage{fancyref}
\usepackage{hyperref}
\usepackage{dcolumn}% align table columns on decimal point
\usepackage{bm}% bold math
\usepackage{tikz}
\usepackage{pgfplots}
\pgfplotsset{compat=1.18}
\tikzset{>=latex}
\usepackage{amsfonts}% math symbols

\usepackage{bibunits}

\usepackage{booktabs}

\usepackage[mode=buildnew]{standalone}% required for including my tikzpictures

\usepackage{lipsum}
\usepackage{graphicx} % Required for inserting images
\usepackage{standalone}% Required for inserting tikz
\usepackage[caption=false]{subfig}% Required for subfigures
\usepackage[percent]{overpic}% Required for writing over pictures
\usepackage{verbatim}% Required for comments
\usepackage{tcolorbox}

\usepackage{physics} % Used in methods relating to LMG
\usepackage{enumerate}
\usepackage{dsfont}
\usepackage{xcolor}

\usepackage[normalem]{ulem}
\usepackage[capitalise]{cleveref}

\usepackage{lipsum}

\begin{document}

\title{Low-depth quantum error correction via three-qubit gates in Rydberg atom arrays}

\author{Laura Pecorari}
\author{Sven Jandura}
\author{Guido Pupillo}

\affiliation{University of Strasbourg and CNRS, CESQ and ISIS (UMR 7006), aQCess, 67000 Strasbourg, France}

\date{\today}

\begin{abstract}
    Quantum error correction (QEC) requires the execution of deep quantum circuits with large numbers of physical qubits to protect information against errors. Designing protocols that can reduce gate and space-time overheads of QEC is therefore crucial to enable more efficient QEC in near-term experiments. Multiqubit gates offer a natural path towards fast and low-depth stabilizer measurement circuits. However, they typically introduce high-weight correlated errors that can degrade the circuit-level distance, leading to slower scalings of the logical error probabilities. In this work, we show how to realize fast and efficient surface code stabilizer readout using only two singly-controlled $Z$ gates acting simultaneously on two target qubits, i.e. two $CZ_2$ gates -- instead of four $CZ$. We show that this scheme is fault-tolerant, and provide a blueprint for implementation in Rydberg atom arrays. We derive the time-optimal pulses implementing the gates and perform extensive QEC numerical simulations with Rydberg decay errors. Compared to the standard protocol using four $CZ$ gates, our scheme is faster, uses fewer gates and crucially maintains fault tolerance with comparable logical error probabilities. 
    Fault-tolerant generalizations of this scheme to biased and erasure-dominant noise models, as well as to other QEC codes, such as quantum Low-Density Parity-Check codes, are also discussed.
\end{abstract}

\maketitle

Quantum error correction (QEC) redundantly encodes the state of single logical qubits into the state of several physical qubits to protect information against errors when the error probability is below a threshold value. Fault tolerance requires spatial error correlations to be as small as possible to guarantee a benign logical error scaling, which is generally achieved by executing pairwise transversal operations between qubits \cite{gottesman1997stabilizercodesquantumerror}. This significantly increases the gate cost for the quantum circuits that implement error correcting codes and realize logical operations with them. At the same time, hardware-specific limitations can prevent the maximum parallelism that is achievable in theory, thus further increasing the circuit depth in practice. For example, in Rydberg atom arrays \cite{saffman_review,morgado,Bluvstein_2023,bluvstein2025architecturalmechanismsuniversalfaulttolerant} the same physical mechanism that enables the gates -- the \emph{Rydberg blockade} -- also prevents the execution of gates acting on neighboring pairs of qubits in parallel, due to cross-talks.
This results in huge gate and space-time overheads that severely constrain experiments and ultimately challenge the path towards error-corrected quantum computation. 

Multiqubit gates provide a natural pathway to reduce the time and gate overhead of QEC and are naturally available in several qubit platforms, such as neutral atoms \cite{PhysRevLett.123.170503,Cao_2024,Yelin_2024}. However, multiqubit gates can introduce high-weight correlated errors that can degrade the circuit-level distance for QEC and lead to a slower scaling of the logical error probability, thus losing fault tolerance. For this reason, fault-tolerant QEC protocols typically rely solely on two-qubit entangling gates.

In this work, we show that it is possible to use multiqubit gates -- specifically, singly-controlled $Z$ gates acting on two target qubits simultaneously ($CZZ$, hereafter shortly denoted as $CZ_2$) -- to efficiently compile the stabilizer measurement circuits of some quantum error correcting codes while preserving fault tolerance, and provide a blueprint for implementation in neutral atom arrays. We design the time-optimal pulses that implement a $CZ_2$ gate with Rydberg atoms and simulate the QEC performance of the surface code \cite{Kitaev_2003,Dennis_2002} when its stabilizers are read out via two $CZ_2$ gates in the presence of Rydberg decay errors. We show that QEC using $CZ_2$ gates uses fewer gates, and is approximately $\times1/3$ faster than existing protocols, while crucially maintaining fault tolerance with comparable logical error probabilities. Our work therefore offers a practical pathway towards efficient and low-depth experimental QEC in Rydberg atom arrays.

We consider quantum error correcting codes implemented with the Steane method for QEC \cite{Steane1997}, where $N$-qubit stabilizer states are prepared by coupling $N$ data qubits, $D_i$, $i\in\{0,1,\dots N\}$, to a common \emph{ancilla} qubit, $A$, such that 
\begin{equation}
    \label{eq:stabilizer}
    |A,D_1,\dots,D_N\rangle\rightarrow|A\oplus D_1\oplus \dots \oplus D_N, D_1,\dots,D_N\rangle
\end{equation}
and their value is read out by measuring the ancilla qubit. Sign flips in the stabilizer eigenvalues signal the possible occurrence of errors.  

In the surface code, $N=4$ and the above transformation is usually performed via four entangling $CZ$ gates to ensure fault tolerance, and repeated at least $D$ times to ensure robustness against measurement errors. This guarantees that $\lceil D/2\rceil$ independent single-qubit errors are necessary to trigger a logical error, corresponding to a $p^{\lceil D/2\rceil}$ scaling of the logical error probability with the physical error probability $p$. If instead the same transformation is performed via one multiqubit $CZ_4$ gate, errors can spread over the four data qubits in all possible ways. Notably, pairs of them can now align with the logical operators so that only $\lceil D/4\rceil$ errors suffice to cause a logical error and the logical error probability scales as $p^{\lceil D/4\rceil}$ \cite{aliferis2005quantumaccuracythresholdconcatenated}. When this is the case, although the logical error probabilities still decay to zero asymptotically, the protocol is often referred to as \emph{non-fault-tolerant}.

These considerations suggest that it should be possible to fault-tolerantly measure stabilizers using two $CZ_2$ gates, provided that pairs of correlated errors do not align to degrade the circuit-level distance of the code. We find that this is indeed the case in the \emph{unrotated} surface code for a specific qubit ordering prescription, while it is never the case in the \emph{rotated} surface code for any qubit ordering prescription, as we explain below.

The unrotated surface code [Fig.~\ref{fig:fig1}(a)] \cite{fowler} encodes one logical qubit in $(D-1)^2+D^2$ physical qubits, that are entangled with $(D-1)^2+D^2-1$ auxiliary ancilla qubits, being $D$ the code distance. The state of the logical qubits is manipulated with the $X_L$ and $Z_L$ logical operators, that are $D$-qubit long Pauli strings connecting opposite array boundaries. With the convention chosen in Fig.~\ref{fig:fig1}(a), $X_L$ is vertical, while $Z_L$ is horizontal. Therefore, when a $Z$ ($X$) stabilizer is affected by errors, pairs of $X$ ($Z$) errors propagated by a single failure should never align vertically (horizontally) to ensure fault tolerance.

We consider a $Z_aZ_bZ_cZ_d$ stabilizer with the qubit ordering specified in Fig.~\ref{fig:fig1}(b) and (c) (the discussion is symmetric for $X$ stabilizers, that only differ by a Hadamard rotation on all the data qubits). In the unrotated surface code, $Z$-stabilizers share two data qubits only with the $X_L$ logical operator. Additionally, CZ gates can only propagate $X$ errors on the control qubit into $Z$ errors on the target qubit. Therefore, in the stabilizer readout scheme via four $CZ$ gates [Fig.~\ref{fig:fig1}(b)], pairs of $Z$ errors caused by an $X$ error on the ancilla atom halfway through the circuit (\emph{hook errors}) never align with the $Z_L$ logical operator. This property ensures that the unrotated surface code is fault-tolerant and intrinsically robust against hook errors \cite{Dennis_2002} for any choice of qubit ordering that preserves the stabilizer commutativity.

In the readout scheme via two $CZ_2$ gates [Fig.~\ref{fig:fig1}(c)], we find that the only ordering that preserves fault tolerance is the one that uses $CZ_aZ_d$ and $CZ_bZ_c$, up to symmetric permutations. In fact, this ordering ensures that pairs of correlated errors introduced by a single failure in any of the two three-qubit gates never align with the corresponding logical operator. Instead, the ordering  $\{CZ_aZ_b,\,CZ_cZ_d\}$ is non-fault-tolerant because single failures during any three-qubit gate can lead to any pair of errors in $\{X,Z,Y\}^{\otimes2}$ aligning with the logical operators. Robustness against hook errors is guaranteed by the topological structure of the code and follows from very similar considerations as above. In the following, we assume that $X$ and $Z$ stabilizers cannot be measured in parallel, as it is typical of neutral atom experiments due to cross-talks and global addressing requirements. Since this trivially ensures stabilizer commutativity, the same qubit ordering is chosen for both $X$ and $Z$ stabilizers [Fig.~\ref{fig:fig1}(c)].

The rotated surface code \cite{bombin} (not shown) encodes one logical qubit in $D^2$ physical qubits using $D^2-1$ ancilla qubits and therefore offers a moderately lower qubit overhead than its unrotated counterpart. However, even when implemented with four $CZ$ gates, it is potentially prone to hook errors, unless its stabilizers are measured following a specific gate ordering prescription \cite{tomita}. That is because now any stabilizer -- either of $X$- or $Z$-type -- shares two data qubits with both the $Z_L$ and $X_L$ logical operators. Due to the impossibility of protecting the stabilizer readout circuit of a rotated surface code against both hook errors and pairs of correlated errors introduced by a $CZ_2$ gate, we find that no fault-tolerant implementation via three-qubit gates exists in this case. A possible exception is the case of a strongly biased noise model (including biased erasure errors). In this case, a fault-tolerant readout scheme via two $CZ_2$ gates can also be found for the rotated surface code. We discuss this case for neutral atom experiments in the Supplementary Material.

\begin{figure}[t]
    \centering
    \includegraphics[scale=1.5]{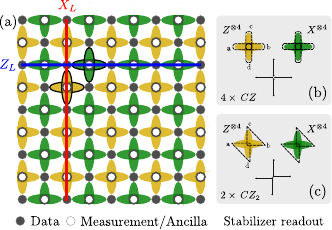}
    \caption{(a) Distance-$5$ unrotated surface code with $X$ (green) and $Z$ (yellow) stabilizers, $Z_L$ (blue) and $X_L$ (red) logical operators are drawn. (b) Fault-tolerant stabilizer readout circuit via four $CZ$ gates. (c) Fault-tolerant stabilizer readout circuit via two $CZ_2$ gates.
    }
    \label{fig:fig1}
\end{figure}

\begin{figure*}[t]
    \centering
    \includegraphics[scale=1.0]{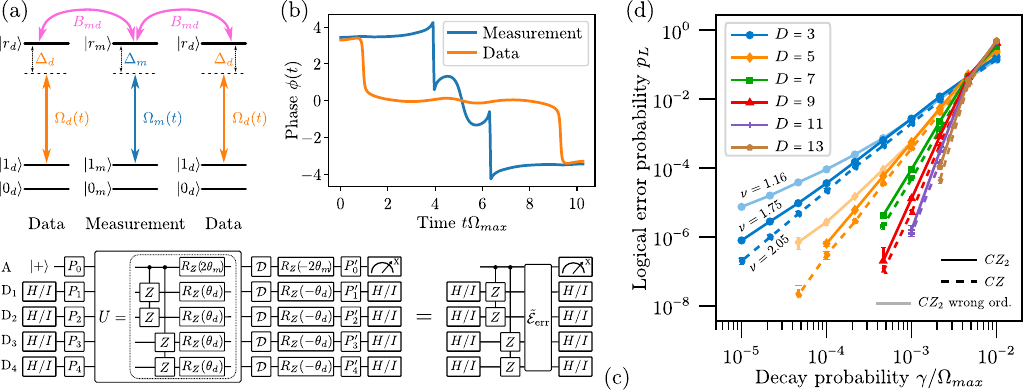}
    \caption{(a) Level diagrams for one measurement ($m$) atom and two data ($d$) atoms, each of them modeled as a three-level system. 
    Only measurement-data Rydberg blockade is assumed, as it can be realized in dual-species experiments. Two different global lasers are used to implement a $CZ_2$ gate with Rabi frequency $\Omega_a(t)$ and detuning $\Delta_a$, $a=m,d$. (b) Time-optimal pulse implementing a $CZ_2$ gates between the measurement (control) and the two data atoms (targets). (c) $X$/$Z$ stabilizer readout circuit using two $CZ_2$ gates with Pauli twirling. (d) QEC performance for the unrotated surface code with time-optimal $CZ_2$ gates and correct qubit ordering (dark solid lines), with time-optimal $CZ_2$ gates and non-fault-tolerant qubit ordering (pale solid lines), and time-optimal $CZ$ gates (dashed lines). The logical error probability is evaluated as the ratio between the number of errors and the total number of shots, performing $D$ rounds of the syndrome extraction circuits, error bars are standard deviations from the Monte Carlo simulations.}
    \label{fig:fig2}
\end{figure*}

We now provide a theoretical blueprint to implement fast surface code stabilizer readout via three-qubit gate in present neutral atoms experiments. 

We start by designing a protocol to realize a high-fidelity and time-optimal $CZ_2$ gate on three atoms. In fact, short gate durations naturally minimize errors arising from the finite lifetimes of the Rydberg states, i.e. spontaneous emission \cite{Jandura_2022}. We refer to the middle atom as \emph{measurement} atom ($m$) and to the other two as \emph{data} atoms ($d$). We model each of them as a three-level system, with long-lived computational states $|0\rangle_a$ and $|1\rangle_a$, and a short-lived auxiliary Rydberg state $|r\rangle_a$, for $a=m,d$. We assume that the atoms interact only via data-measurement Rydberg blockade. This setup can naturally be realized in dual-species neutral atom experiments, which allow for separate control of inter-species and intra-species interactions \cite{PhysRevA.92.042710,Singh_2023,Bernien,Ireland_2024}. For example, this protocol could be used to implement a $CZ_2$ gate in Rb-Cs dual-species arrays, where inter-species entangling gates are realized via F\"orster resonance-mediated dipolar interaction enabled by the nearly degenerate pairs of states $|68S_{1/2}\rangle_\text{Rb}$, $|67S_{1/2}\rangle_\text{Cs}$ and $|68P_{1/2}\rangle_\text{Rb}$, $|67P_{3/2}\rangle_\text{Cs}$ \cite{PhysRevA.92.042710,Bernien}.
Additionally, the use of two different atomic species naturally breaks the permutation symmetry between control and target qubits required by single-control multi-target gates. A similar protocol in single-species arrays would instead require extra atomic levels, or local controls, or spatial Rydberg blockade engineering \cite{stein2025multitargetrydberggatesspatial}.

We look for a laser pulse acting simultaneously on the data atoms and a different laser pulse acting on the measurement atom, so that the system follows the evolution
\begin{equation}
    \label{eq:gate}
    |md_1d_2\rangle\rightarrow\text{exp}[i\theta_mm+i(\theta_d+m\pi)(d_1+d_2)]|md_1d_2\rangle,    
\end{equation} 
where $m\in\{0,1\}$ and $d_j\in\{0,1\}_{j=1,2}$ denote the initial states of the measurement and data atoms. This evolution is equivalent to two simultaneous $CZ$ gates between the measurement and each data atom up to a global $Z(\theta_m)$ and $Z(\theta_d)$ single-qubit rotation on the measurement and data atoms, respectively. We assume infinite blockade strength between the measurement atom and each data atom, and write the single-atom Hamiltonians as $H_a=(\Omega_a(t)/2)|r_a\rangle\langle1_a|+\text{h.c.}+\Delta_a(t)|r_a\rangle\langle r_a|$ for $a=m,d$, where $\Omega_a(t)=|\Omega_a(t)|e^{i\phi_a(t)}$ is the complex Rabi frequency of the incident laser coupling $|1_a\rangle\leftrightarrow|r_a\rangle$ with amplitude $|\Omega_a(t)|$ and phase $\phi_a(t)$, and $\Delta_a(t)=d\phi_a(t)/dt$ is the detuning [see Fig.~\ref{fig:fig2}(a)]. The total three-atom Hamiltonian then reads $H=PH'P$, where 
\begin{equation*}
H'=H_m\otimes I_d\otimes I_d+I_m\otimes H_d\otimes I_d+I_m\otimes I_d\otimes H_d
\end{equation*}
and $P$ is the projector onto the non-blockaded states, 
\begin{align*}
\begin{split}
    P=I_m\otimes I_d\otimes I_d&-|r_m\rangle\langle r_m|\otimes|r_d\rangle\langle r_d|\otimes I_d \\
    &-|r_m\rangle\langle r_m|\otimes I_d\otimes|r_d\rangle\langle r_d|\\
    &+|r_m\rangle\langle r_m|\otimes|r_d\rangle\langle r_d|\otimes|r_d\rangle\langle r_d|.
\end{split}
\end{align*}

We note that two data atoms can be simultaneously Rydberg-excited, as we only assume measurement-data blockade.

Following Ref.~\cite{Jandura_2022}, we use the GRAPE algorithm to find the shortest pulses implementing the evolution in Eq.~\eqref{eq:gate}, which we plot for the phases in Fig.~\ref{fig:fig2}(b). We find that the time-optimal pulses have amplitudes equal to the maximal Rabi frequency, $|\Omega_a(t)|=\Omega_{max}$ for $a=m,d$ (not shown). The optimal pulse duration is $T\Omega_{max}=10.27$ and the infidelity is $1-\mathcal{F}\approx3.90\times10^{-8}$. We also find that the measurement atom spends $T_{Ry}\Omega_{max}\approx3.36$ in the Rydberg state, whereas for the data atoms this time is $T_{Ry}\Omega_{max}\approx1.91$. We observe that, compared to a three-qubit $\pi$-$2\pi$-$\pi$ pulse, the time-optimal pulse has a smooth shape, which simplifies experimental implementation. Interestingly, it is approximately $18\%$ faster, the time the measurement atom spends in the Rydberg state is decreased by $29\%$, while that of the data atoms is increased by $21\%$, and the overall time spent in the Rydberg state by the three atoms is decreased by $9\%$. Thus, this pulse is less prone to errors resulting from spontaneous emission. Additionally, the difference in the time spent in the Rydberg states of measurement and data atoms suggests that experiments could further benefit from pulses optimizing these times  separately, rather than the total pulse duration, depending on the specific experimental implementation.  
Finally, we observe that both pulses acting on measurement and data atoms in Fig.~\ref{fig:fig2}(b), while smooth, show two phase jumps of approximately $\sim\pi$. However, these can be removed by optimizing the pulse for the detuning -- instead of the phase -- and fixing a cutoff $|\Delta_a|<\Delta_{max}$, with minimal loss of pulse performance (see Supplementary Material).

We then use \texttt{Stim} \cite{gidney2021stim} and the methodology developed in Ref.~\cite{jandura2024surfacecodestabilizermeasurements} to simulate the QEC performance of unrotated surface codes in the presence of Rydberg decay errors when their stabilizers are read out with the $CZ_2$ gates modeled above. 
The choice of accounting for only Rydberg decay errors is motivated by Rydberg leakages being the most detrimental error mechanism affecting neutral atom qubits. In fact, they are found to dominate the infidelity budget by more than one order of magnitude compared to other error sources \cite{senoo2025highfidelityentanglementcoherentmultiqubit}. Therefore, our noise mode is physically motivated, e.g., in alkali atoms, where the computational subspace is encoded in two hyperfine states of the ground state manifold of the atom. Leakages that do not result in atom loss (for which different gadgets are necessary to ensure fault tolerance \cite{perrin2025quantumerrorcorrectionresilient, baranes2025leveragingatomlosserrors}), will result in decays to the ground state manifold, either directly or via different intermediate Rydberg states (e.g., by blackbody radiation-induced transitions, finite Rydberg lifetime, under- or over-rotation during the gate, etc.). Decays to the wrong hyperfine ground states outside the qubit subspace can instead be mitigated via optical pumping \cite{PhysRevX.12.021049}.

Thus, $X/Z$ stabilizers are read out by performing $D$ rounds of the circuit shown in Fig.~\ref{fig:fig2}(c): We first reset the ancilla qubit to $|+\rangle=(|0\rangle+|1\rangle)/\sqrt{2}$ and apply a $5$-qubit unitary $U$ consisting of two three-qubit gates applied with the ordering prescription identified above. After each gate, if an error has occurred, we let the system relax back to the computational space. The unitary $U$ is chosen in such a way that in the absence of Rydberg state decay ($\gamma=0$), it exactly implements the evolution shown in Fig.~\ref{fig:fig2}(b). Instead, to treat the case $\gamma>0$, after the perfect evolution, we apply an amplitude damping with probability $\gamma$ assuming a branching ratio $1$:$1$ for the two ground state levels $|0\rangle$ and $|1\rangle$. Finally, we apply an amplitude damping channel on each qubit with probability $1$ to remove all the Rydberg population left after $U$, $\mathcal{D}(\rho)=\Pi\rho\Pi+\langle r|\rho|r\rangle\Pi/2$ with $\Pi=|0\rangle\langle0|+|1\rangle\langle1|$, and correct for the single-qubit rotations, $R_Z(\theta_a)$, $a=m,d$, to ensure that the final evolution is equivalent to two $CZ_2$ gates. This can be experimentally achieved by either waiting long enough for the Rydberg state to decay back to the ground state or by coupling the Rydberg state to a short-lived intermediate state decaying to $|0\rangle$ or $|1\rangle$ \cite{PhysRevX.12.021049}. Rydberg decays propagating between different stabilizer readout rounds are not captured with this model. This is consistent with optical traps typically being designed to be anti-trapping for atoms in the Rydberg state, hence Rydberg excitations at the end of a readout round more likely result in atom loss than drift \cite{PhysRevA.108.023122}.

A similar circuit is constructed to simulate the readout via four $CZ$ gates with the standard time-optimal pulses \cite{Jandura_2022}. We note that in~\cite{jandura2024surfacecodestabilizermeasurements} it was found that the time-optimal $CZ$ gate does not preserve fault tolerance in the rotated surface code because pairs of hook $Z$ errors can align vertically, horizontally, and diagonally, which motivated the search for  different pulses. However, this is not an issue in the unrotated surface code for the reasons explained above.

We also observe that in the transformation in Eq.~\eqref{eq:gate}, a missing data atom behaves as a qubit in state $|0\rangle$, hence the pulse designed for $CZ_2$ gates implements  a $CZ$ gate -- albeit not time-optimally -- when shined on only one measurement and one data atom. For this reason, our protocol does not require special addressability for boundary stabilizers, where only three data qubits are present instead of four, which we have modeled numerically (see more comments in the Supplementary Material).

This stabilizer readout circuit is then equivalent to two ideal $CZ_2$ gates followed by a $5$-qubit noise channel $\tilde{\mathcal{E}}_{err}(\rho)$, which can be simulated as a Clifford channel via randomized compiling by inserting random single-qubit Pauli gates at the beginning ($P$) and at the end ($P'$) of the circuit (or without, corresponding to the Pauli twirling approximation) \cite{PhysRevA.88.012314}. Thus, $\tilde{\mathcal{E}}_{err}(\rho)=\sum_Q\lambda_QQ\rho Q$, with $Q$ any $5$-qubit Pauli string and $\lambda_Q$ its probability of occurring (see Supplementary Material).

We show the results of our QEC numerical simulations in Fig.~\ref{fig:fig2}(d). Here we compare the QEC performance of unrotated surface codes implemented with two $CZ_2$ gates and the correct qubit ordering (dark solid lines), with two $CZ_2$ gates and the wrong qubit ordering (pale solid lines), and with four $CZ$ gates (dashed lines). In all cases, error syndromes have been decoded with \texttt{pymatching} \cite{Higgott2025sparseblossom}. For distance-$3$ surface codes, the last five points are fitted with a $\sim\gamma^\nu$ functional dependence to extract the minimum number of uncorrectable errors, which is $\lceil D/2\rceil$ for fault-tolerant implementations, and $\lceil D/4\rceil$ otherwise. The $4\times CZ$ scheme is fault-tolerant and $\nu\approx2$ with good approximation. For the wrong qubit ordering, as expected, the $2\times CZ_2$ scheme is non-fault-tolerant and one error suffices to create a logical error, $\nu\approx1$, in a distance-$3$ code. When enforcing the correct qubit ordering, the $2\times CZ_2$ scheme ($D=3$) is fault-tolerant with a $\nu\approx2$. The $\nu_{CZ_2}<\nu_{CZ}$ is likely due to a suboptimal error decomposition in the numerical simulations \cite{old2025faulttolerantstabilizermeasurementssurface}, which could be mitigated by using a non-matching based decoder. However, since the higher decoding accuracy typically comes at the price of lower decoding speed, we believe that \texttt{pymatching} still represents the best decoding strategy given the physical error probabilities affecting today's neutral atom experiments, $p>10^{-4}$. We also note that for the higher distance codes we have considered in this work, no degradation of the scaling is observed in the scanned regime of decay probabilities.

These results show that the fault-tolerant implementation via $CZ_2$ gates offers only slightly higher logical error probabilities compared to the implementation via four $CZ$ gates. That is because three-qubit gates spread more errors than two-qubit gates (e.g., the decay of one atom during the gate is more likely to introduce errors also on the other atoms), hence increasing the noise entropy, while still preserving fault tolerance. However, the round-time of a single stabilizer (up to single-qubit gates and measurements) is approximately $\times1/3$ faster ($2\times T_{CZ_2}\Omega_{max}=20.54$ and $4\times T_{CZ}\Omega_{max}=30.44$). 

In conclusion, we have proposed a new scheme for the implementation of surface code quantum memories in today's Rydberg atom experiments using two $CZ_2$ gates in a fault-tolerant manner. Crucially, our fault-tolerant scheme is $\times1/3$ faster, uses half the number of gates, and offers comparable logical error probabilities compared to the standard scheme using four $CZ$ gates.

Although in this work we have focused on the surface code for contact with the experiments, we note that our scheme can be generalized to other QEC codes. This could be particularly relevant for high-rate quantum Low-Density Parity-Check (qLDPC) codes \cite{PRXQuantum.2.040101,Bravyi_2024,xu2023constantoverheadfaulttolerantquantumcomputation,Pecorari_2025,poole2025,saffman2025quantumcomputingatomicqubit}, which enjoy appealing encoding properties, but typically require larger compilation circuits due to the higher weight of their stabilizers. For example, La-cross qLDPC codes \cite{Pecorari_2025} could be fault-tolerantly implemented via three $CZ_2$ gates -- one targeting the two distant qubits and the other two as outlined above for the surface code -- instead of six $CZ$ gates in unrotated arrays with open boundaries (see Supplementary Material). It remains an open question whether other qLDPC codes could benefit from similar fault-tolerant implementations via multiqubit gates.

Erasure conversion \cite{Wu_2022,Sahay_2023,Ma_2023,pecorari2025quantumldpccodeserasurebiased,zhang2025leveragingerasureerrorslogical} and loss-detection protocols \cite{Chow_2024,perrin2025quantumerrorcorrectionresilient,baranes2025leveragingatomlosserrors} that require interleaving the stabilizer readout circuits with additional operations to detect specific errors could also take significant advantage from the lower depth of our scheme (see Supplementary Material). 

Moreover, although in this work we have not accounted for calibration errors (Doppler shift, laser fluctuations, etc.) \cite{PRXQuantumJandura}, whose strength can increase with the qubit number, it is reasonable expectation that these errors will also benefit from the higher speed of the protocol, thus compensating for the larger infidelity contributions. 

Finally, this new readout scheme can also be generalized to, e.g., trapped ions and superconducting qubits, where multiqubit gates are natively available and other error mechanisms are dominant. Therefore, it is an open question whether this scheme could pave the way for improvements in threshold and logical error probabilities, along with cycle time, in these platforms.
 
\emph{Data of the QEC simulations and the time-optimal pulses shown in the main text are available at \cite{Pecorari2025data}.}

\emph{We gratefully acknowledge discussions with Hannes Bernien on $CZ_2$ gates for dual-species Rydberg atom arrays. We also thank Shannon Whitlock for fruitful discussions and comments on the manuscript. This research has received funding from the European Union’s Horizon 2020 research and innovation programme under the Horizon Europe programme HORIZON-CL4-2021-DIGITAL-EMERGING-01-30 via the project 101070144 (EuRyQa) and from the French National Research Agency under the Investments of the Future Program projects ANR-21-ESRE-0032 (aQCess), ANR-22-CE47-0013-02 (CLIMAQS), ANR-17-EURE-0024 (QMat), and ANR-22-CMAS-0001 France 2030 (QuanTEdu-France).}

\emph{While completing this work, we became aware of related work on quantum error correction via three-qubit gates with surface codes in Ref.~\cite{old2025faulttolerantstabilizermeasurementssurface} and on optimizing three-qubit gates with superconducting transmon qubits for surface-code-based quantum error correction in Ref.~\cite{tasler2025optimizingsuperconductingthreequbitgates}.}

\bibliography{references}

%%%%%%%%%% Merge with supplemental materials %%%%%%%%%%
\onecolumngrid
\clearpage
\begin{center}
\textbf{\large Supplementary Material for ``Low-depth quantum error correction via three-qubit gates in Rydberg atom arrays"}

\vspace{0.5cm}

Laura Pecorari, Sven Jandura, and Guido Pupillo

\emph{University of Strasbourg and CNRS, CESQ and ISIS (UMR 7006), aQCess, 67000 Strasbourg, France}

{\small (Dated: \today)}
\end{center}
\maketitle
%%%%%%%%%% Merge with supplemental materials %%%%%%%%%%
%%%%%%%%%% Prefix a "S" to all equations, figures, tables and reset the counter %%%%%%%%%%
\setcounter{equation}{0}
\setcounter{figure}{0}
\setcounter{table}{0}
\makeatletter
\renewcommand{\thesection}{S\arabic{section}}
\renewcommand{\theequation}{S\arabic{equation}}
\renewcommand{\thefigure}{S\arabic{figure}}
%%%%%%%%%% Prefix a "S" to all equations, figures, tables and reset the counter %%%%%%%%%%
\twocolumngrid

\section{Rotated surface code with biased noise}
\begin{figure}[ht]
    \centering
    \includegraphics[scale=2.0]{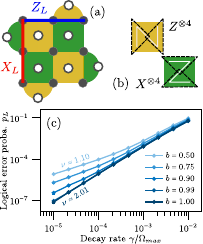}
    \caption{(a) Distance-$3$ rotated surface code, $Z_L$ (blue) and $X_L$ (red) logical operators are shown. (b) $Z$- (yellow) and $X$-type stabilizers with fault-tolerant $CZ$ gate ordering (white) and chosen $CZ_2$ gate ordering (dashed black). (c) Logical error rate of a distance-$3$ rotated surface code for different branching ratios, showing that for purely $Z$-biased noise ($b=1.00$) the three-qubit gate scheme is fault-tolerant. Error bars are within the marker sizes.}.
    \label{fig:figBIAS}
\end{figure}

In this supplementary section we discuss the case of a noise model strongly biased towards $Z$ errors, which allows to partially recover fault tolerance on the rotated surface code [Fig.~\ref{fig:figBIAS}(a)] when its stabilizers are read out using two $CZ_2$ gates. 

We have discussed in the main text that for generically unbiased noise, the three-qubit gate readout scheme proposed in this work is not fault-tolerant for the rotated surface code due to the impossibility to simultaneously prevent hook errors and pairs of errors on two data qubits introduced by the three-qubit gates. However, if the gate noise is completely biased towards $Z$ errors, no hook errors can occur and fault tolerance is recovered, provided that the three-qubit gates act on pairs of data qubits that do not align with the $Z_L$ logical operator. A possible pairing choice is shown in Fig.~\ref{fig:figBIAS}(b) (black dashed lines describe the data qubit pairing for the $CZ_2$ scheme, while white smooth lines describe the standard fault-tolerant $CZ$ gate ordering for the rotated surface code).

A noise model completely biased towards $Z$ errors corresponds to a branching ratio $b=1.0$, that is, the whole Rydberg population always decays back to $|1\rangle$ and never to $|0\rangle$. Although it is possible to identify Rydberg states with the desired branching ratio by exploiting selection rules, such a noise model is typically extremely hard to engineer in practice, due to blackbody radiation-induced transitions occurring at room temperature. A possible strategy to inhibit these transitions -- and consequently preserve a strong bias towards $Z$ errors -- is to embed the system in a cryogenic environment \cite{PhysRevApplied.22.024073,PRXQuantum.6.020337}. Another possibility is to convert large fractions of decays into biased erasure errors (see below) via erasure conversion protocols in Alkline-earth(-like) atom platforms. In other qubit architectures, such as trapped ions and superconducting qubits, strong noise biases might instead be easier to engineer \cite{Puri_2020}.

In order to assess how strong the noise bias has to be for the scheme to be fault-tolerant with good approximation, we plot in Fig.~\ref{fig:figBIAS}(c) the logical error rate of a distance-$3$ rotated surface code for different branching ratios, corresponding to different bias strengths. For $b=1.00$ (no differences are observed for $b=0.99$), the three-qubit gate readout scheme is fault-tolerant, with $\nu=\lceil D/2\rceil\approx2.01$ for $p_L\propto \gamma^\nu$. Instead, when the branching ratio is decreased from $b=0.90$ to $b=0.50$, the logical error probabilities increase due to the larger noise entropy (i.e. the larger number of realized error combinations), and fault tolerance is gradually lost. In fact, by fitting the last five points of the decoding curves, we find slopes: $\nu_{b=0.90}\approx1.68$, $\nu_{b=0.75}\approx1.30$, $\nu_{b=0.50}\approx1.10$. These results suggest that a noise bias at least as strong as for $b=0.90$ is necessary for the three-qubit gate scheme to be approximately fault-tolerant in the experimentally relevant regime $p\sim10^{-4}-10^{-3}$.

\section{La-cross qLDPC codes}
High-rate quantum Low-Density Parity-Check (qLDPC) codes offer larger encoding rates than the surface code, at the price of higher-weight stabilizers and long-range connectivity that make their implementation in near-term quantum devices challenging \cite{saffman2025quantumcomputingatomicqubit}. These codes also suffer a poorer parallelism compared to the surface codes, both due to the larger number of gates they require and to hardware-connectivity limitations. In neutral atom arrays, it has been proposed to implement the necessary long-range connectivity either via atom rearrangement \cite{xu2023constantoverheadfaulttolerantquantumcomputation} or via static long-range interactions enabled by high principal quantum number Rydberg states \cite{Pecorari_2025,poole2025}. The former scheme requires large numbers of movements, the latter is severely limited by Rydberg blockade constraints.

In this section, we comment on how one could fault-tolerantly implement qLDPC codes via low-depth quantum circuits using multiqubit gates, generalizing what was discussed in the main text for the surface code.
We focus on La-cross qLDPC codes, which are hypergraph product codes built from equal classical cyclic codes, similarly to the surface code \cite{Pecorari_2025}. La-cross codes have weight-$6$ stabilizers acting on four neighboring and two far apart data qubits. Three qubits in the support of the stabilizers are vertical and three horizontal, and logical operators are vertical/horizontal Pauli strings. When implemented in a unrotated array with open boundaries, La-cross codes have proven to be robust against hook errors as the unrotated surface code \cite{Pecorari_2025}. We also note that, since by definition the logical operators commute with the stabilizers and they are all either vertical or horizontal, it follows that each stabilizer shares at most two qubits with a logical operator, as for the surface code. Hence, the worst-case scaling for the logical error probability is still $p^{\lceil D/4\rceil}$ due to pairs of correlated $Z$ ($X$) errors aligning with the $Z_L$ ($X_L$) logical operators. This could be avoided by using three $CZ_2$ gates, one for the two distant qubits and two for the four neighboring qubits, as in the unrotated surface code discussed in the main text. The $CZ_2$ targeting the two distant data qubits can be implemented either by moving the data atoms closer to the ancilla or statically by identifying resonant Rydberg states with higher principal quantum numbers.

\section{Optimal pulses without phase jumps}
In the main text and in Fig.~\ref{fig:fig2}(b), we have discussed the time-optimal pulse implementing a $CZ_2$ gate between one measurement atom and two data atoms assuming only measurement-data Rydberg blockade. This pulse has total duration $T\Omega_{max}=10.27$, infidelity $1-\mathcal{F}\approx3.90\times10^{-8}$ and the time spent in the Rydberg state is $T_{Ry}\Omega_{max}\approx3.36$ and $T_{Ry}\Omega_{max}\approx1.91$ for the measurement and data atoms, respectively. Such time-optimal pulse displays two phase jumps of approximately $\sim\pi$ for both measurement and data atoms. In this section, we show that they can be eliminated with minimal loss of pulse performance.

\begin{figure}[htb]
    \centering
    \includegraphics[scale=0.98]{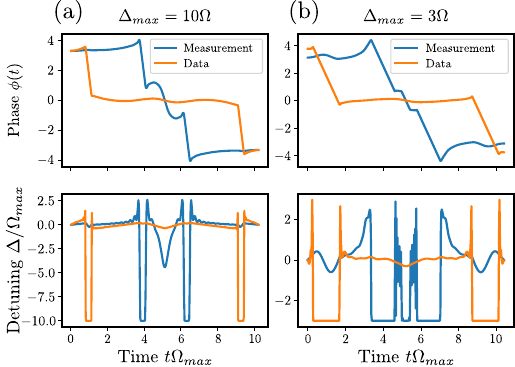}
    \caption{Pulses implementing a $CZ_2$ gate between one measurement atom and two data atoms assuming only measurement-data Rydberg blockade with different cutoffs on the maximum detuning $\Delta_{max}$.}.
    \label{fig:figs1}
\end{figure}

We now parametrize the pulse in terms of amplitude and detuning -- instead of phase -- and fix a maximum cutoff $|\Delta_a|\leq M\Omega_a$, for $a=m,d$. We show in Fig.~\ref{fig:figs1} the resulting pulses for $M=10$ and $M=3$, respectively. For $\Delta_{max}=10\Omega_a$ [Fig.~\ref{fig:figs1}(a)], the total pulse duration is $T\Omega_{max}=10.25$, infidelity $1-\mathcal{F}\approx3.87\times10^{-6}$ and the time spent in the Rydberg state is $T_{Ry}\Omega_{max}\approx3.34$ and $T_{Ry}\Omega_{max}\approx1.90$ for the measurement and data atoms, respectively. The phase jumps are no longer present in the phase, but there is now a phase jump in the detuning. We can eliminate such phase jump by lowering the maximum cutoff, for example $\Delta_{max}=3\Omega_a$ [Fig.~\ref{fig:figs1}(b)], for which the total pulse duration is $T\Omega_{max}=10.40$, infidelity $1-\mathcal{F}\approx3.65\times10^{-7}$ and the time spent in the Rydberg state is $T_{Ry}\Omega_{max}\approx3.38$ and $T_{Ry}\Omega_{max}\approx1.89$ for the measurement and data atoms, respectively. In this case, the detuning shows no phase jumps, but displays some oscillations. Given that no oscillations are visible in the phase and $\Delta(t)=d\phi(t)/dt$, we conclude that these are just numerical instability and are not necessary for the successful implementation of the gate. 

\section{Details on error correction simulations}
In this section, we provide more details on the QEC simulations shown in the main text. The dynamics of a five-qubit surface code stabilizer (assuming only data-ancilla blockade) is governed by the Lindblad master equation $\dot{\rho}=-i[H,\rho]+\sum_{q,i}L_i^{(q)}\rho L_i^{(q)\dagger}-\{L_i^{(q)\dagger}L_i^{(q)},\rho\}$ with Hamiltonian ($\hbar=1$)
\begin{align}
    \label{eq:hamiltonian}
    \begin{split}
        H=\sum_{i=0}^4 &B_{mi}|r_mr_{d,i}\rangle\langle r_mr_{d,i}|+ \\
        &+\sum_{\substack{a=\{m,d\}\\i=\{0,\dots,4\} }}\frac{\Omega_{a,i}(t)}{2}|r_{a,i}\rangle\langle1_{a,i}|+\text{h.c.}
    \end{split}
\end{align}
and collapse operators $L_i^{(q)}=\sqrt{\gamma/2}|q_{a,i}\rangle\langle r_{a,i}|$ with $q\in\{|0_a\rangle,|1_a\rangle\}$, $a\in\{m,d\}$ and $i\in\{0,\dots,4\}$. Thus, we numerically integrate the Lindblad master equation for $H$ and $L_i^{(q)}$ with initial condition $\rho(0)=R$ for any $5$-qubit Pauli string $R$. Then, following Ref.~\cite{jandura2024surfacecodestabilizermeasurements}, since the $\lambda_Q$ are the diagonal coefficients of the $\chi$-matrix representation of the non-twirled error channel $\mathcal{E}_{err}(\rho)=\sum_{Q,Q'}\chi_{Q,Q'}Q\rho Q'$, upon Pauli twirling we find
\begin{equation}
    \label{eq:lambdas}
    \lambda_Q=4^{-5}\sum_Qs(R,Q)\text{tr}\left(R\mathcal{E}_{err}(R)\right)
\end{equation}
being $\mathcal{E}_{err}(R)=U^\dagger\mathcal{D}^{\otimes5}\rho(T)U$ for the $U$ defined in the main text, $T$ the total pulse duration, and $s(R,Q)=\pm1$ if $R$ and $Q$ commute/anticommute.

\section{Boundary contributions}
\begin{figure}[htb]
    \centering
    \includegraphics[scale=0.97]{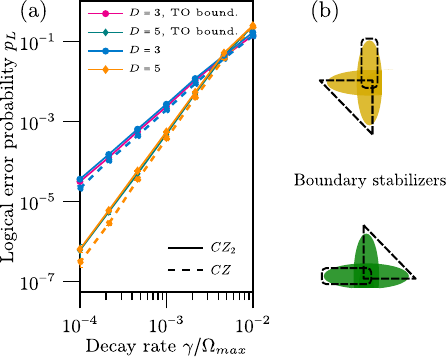}
    \caption{(a) Boundary contributions to the logical error rate for low-distance unrotated surface codes (namely, $D=3,5$). Solid lines correspond to the implementation via $CZ_2$ without special addressability at the boundary, i.e. assume the same global pulse to implement both $CZ_2$ and $CZ$ gates at the boundaries, the latter being non time-optimal. Dashed lines correspond to the implementation via four time-optimal $CZ$ gates. The other solid lines (magenta and teal) are for the $CZ_2$ case with time-optimal $CZ$ at the boundaries (at the expense of losing global addressability across the array). (b) Example of weight-$3$ $Z$ (yellow) and $X$ (green) boundary stabilizers requiring one $CZ_2$ and one $CZ$ gate to be read out.}
    \label{fig:figs2}
\end{figure}

In the main text, we have shown how to readout the stabilizers of unrotated surface codes using two $CZ_2$ gates instead of four $CZ$ gates. However, boundary stabilizer are weight-$3$ operators and they require one $CZ_2$ and one $CZ$ gate [Fig.~\ref{fig:figs2}(b)]. We see that in the evolution in Eq.~\eqref{eq:gate} a missing data atom behaves as an atom in state $|0\rangle$, hence the same pulse can be shined on one measurement and one data atom to implement the necessary $CZ$ gate at the boundary, albeit not time-optimally. Preserving global addressability across the qubit array strongly simplifies the experiments, and therefore in the main text we have used the same pulse for bulk and boundary stabilizers. In this section we quantify the loss of speed, and hence of logical fidelity due to the slower boundary contributions. We show in Fig.~\ref{fig:figs2}(a) QEC simulations for low-distance surface codes -- that are those mostly impacted by boundary contributions -- and compare implementations via four time-optimal $CZ$ gates (dashed lines), via two time-optimal $CZ_2$ with non-time-optimal boundaries (solid lines), and via two time-optimal $CZ_2$ with time-optimal boundaries (solid lines, different colors). The latter shows lower logical error probabilities compared to the global protocol, although still slightly higher than those offered by the standard scheme that uses four $CZ$ gates. For larger-distance codes, the bulk contribution will dominate over the boundary ones. We stress that, while the first two protocols only require global addressability, the last one would require local addressability at the boundaries in realistic implementations.

\vspace{1cm}

\section{Multiqubit readout and erasure conversion}
In this supplementary section we qualitatively discuss how erasure conversion can benefit from the three-qubit gate readout scheme discussed in this work.

In alkaline-earth(-like) atoms, the qubit subspace is typically encoded into long-lived metastable states and gates are performed via an auxiliary Rydberg state accessible via single-photon transition. It has been theoretically predicted that up to $98\%$ of gate errors can be converted into \emph{erasures}, that is decays to the true ground state of the atom that can be detected. Erasure conversion therefore requires interleaving gates with ground state fluorescence imaging to herald error locations without disturbing the metastable states. Atoms involved in the erased gates are then replaced with fresh ones, typically initialized in state $|1\rangle$ \cite{Wu_2022,Sahay_2023,Ma_2023}.

The overhead of erasure conversion is strictly dependent on the depth of the stabilizer readout circuit and, therefore, can be significantly improved by a scheme using fewer multiqubit gates. However, the alkali-inspired noise model that we have enforced in this work is unsuitable for alkaline-earth(-like) atoms, where most of the Rydberg decays occur out of the qubit subspace. This would then require either a more refined modeling to account for both ground and metastable manifolds or a more generic depolarizing noise model. Instead, erasure errors (described as the combination of decay and reloading operation) are typically modeled as two-qubit depolarizing or $Z$-biased Pauli errors with known locations \cite{Wu_2022,Sahay_2023,pecorari2025quantumldpccodeserasurebiased}. The surface code stabilizer readout scheme using two $CZ_2$ gates with three-qubit depolarizing noise has recently been studied in Ref.~\cite{old2025faulttolerantstabilizermeasurementssurface}, showing some improvement over the standard scheme using four $CZ$ gates with two-qubit depolarizing noise. Additionally, if the erased atoms are replaced with fresh ones in state $|1\rangle$, the combination of heralded loss and recovery operation is equivalent -- up to Pauli twirling -- to a $Z$-biased noise model. Therefore, in the limit where most errors are \emph{biased} erasures, the fault tolerance of the three-qubit gate scheme is recovered with good approximation also for the rotated surface code (see first supplementary section). This ultimately suggests that erasure conversion could significantly benefit from our low-depth multiqubit readout scheme, even under the assumption of negligible idle errors. 

\end{document}